\documentclass{article}

\usepackage{amsmath,graphicx,mlspconf}

\usepackage{amssymb}
\usepackage{amsmath}
\usepackage{multicol}
\usepackage{url}
\usepackage{epsfig}
\usepackage{color}
\usepackage[super]{nth}
\usepackage{caption}
\usepackage{setspace}
\usepackage{times}
\usepackage{verbatim}
\usepackage{tabularx}
\usepackage{algorithmic,algorithm2e,float}
\usepackage{bm}
\usepackage{adjustbox}

\DeclareMathOperator*{\argmax}{argmax}
\newcommand*{\argmaxl}{\argmax\limits}
\newcommand{\norm}[1]{\left\lVert #1 \right\rVert}

\makeatletter
\renewcommand{\section}{\@startsection {section}{1}{\z@}%
             {-3.5ex \@plus -1ex \@minus -.2ex}%
             {2.3ex \@plus .2ex}%
             {\normalfont\Large\scshape\bfseries}}
\makeatother

\begin{document}

\title{Efficient Capon-based approach exploiting temporal windowing for electric network frequency estimation}%
\name{George Karantaidis and Constantine Kotropoulos}
\address{Department of Informatics, Aristotle University of Thessaloniki,
Thessaloniki, 54124, Greece}

%\ninept
\maketitle

\begin{abstract}
Electric Network Frequency (ENF) fluctuations constitute a powerful tool in multimedia forensics. An efficient approach for ENF estimation is introduced with temporal windowing based on the filter-bank Capon spectral estimator. A type of Gohberg-Semencul factorization of the model covariance matrix is used due to the Toeplitz structure of the covariance matrix. Moreover, this approach uses, for the first time in the field of ENF, a temporal window, not necessarily the rectangular one, at the stage preceding spectral estimation. Krylov matrices are employed for fast implementation of matrix inversions. The proposed approach outperforms the state-of-the-art methods in ENF estimation, when a short time window of $1$ second is employed in power recordings. In speech recordings, the proposed approach yields highly accurate results with respect to both time complexity and accuracy. Moreover, the impact of different temporal windows is studied. The results show that even the most trivial methods for ENF estimation, such as the Short-Time Fourier Transform, can provide better results than the most recent state-of-the-art methods, when a temporal window is employed. The correlation coefficient is used to measure the ENF estimation accuracy.

\end{abstract}

\section{INTRODUCTION}

Multimedia content is present in all aspects of everyday life, containing citizens' sensitive information. Audio, image, and video recordings are vulnerable to editings, alterations, and all kind of attacks trying to modify their content. It is evident that accurate methods in the field of multimedia forensics are necessary to confront such criminal actions and at the same time to authenticate the content itself.

A ground-breaking tool for multimedia authentication was introduced in  \cite{b1}. The so-called Electric Network Frequency (ENF) criterion takes advantage of the fact that the power grid acts as a digital watermark in multimedia recordings. ENF fluctuations are embedded in audio, images, and video signals. They are generated by the instantaneous difference between the produced and the demanded power supply. The nominal value of ENF is $60 \,Hz$ in the U.S. and $50 \, Hz$ in Europe. Instantaneous value varies in time, depending on consumers' behavior. ENF behaves as a fingerprint for digital recordings due to its properties, as presented in \cite{b1}.

An abundance of ENF studies focus on extracting the ENF signal in order to verify the authenticity, time, and location of recordings \cite{b2}. A non-parametric adaptive method based on dynamic programming is presented in \cite{b3}. It overcomes the interference present within signals and the drawbacks arisen from the periodogram-based methods. An efficient Maximum-Likelihood Estimation approach is presented in \cite{b4}, which takes into account the higher harmonics of the signal in addition to the fundamental frequency. The Cramer-Rao bound is employed in order to point out the best expected performance in ENF extraction. An approach, which combines multiple harmonics of ENF, adopting a local signal-to-noise ratio (SNR) for each harmonic is presented in \cite{b5}. A Discrete Fourier Transform based algorithm is presented in \cite{b6}, where ENF is extracted by specific spectral lines instead of  the conventional entire frequency band. A weighted linear prediction approach followed by a rank reduction method for de-noising power or speech recordings and accurate ENF estimation is introduced in \cite{b7}. A systematic study of parametric and non-parametric methods for ENF extraction is presented in \cite{b8}, where it is shown that fine tuning of signal filtering is of crucial importance for accurate frequency estimation. Moreover, efficient parametrization of extraction methods is found to affect critically ENF estimation.

ENF fluctuations are detected also in video recordings captured in indoor environments where fluorescent light is present. This can lead to accurate timestamp verification and blaze a trail for more real world applications, such as child pornography and terrorism \cite{b12}. The rolling shutter mechanism can be used in order to effectively extract ENF from video recordings \cite{b10}. Recently, ENF fingerprint is demonstrated to exist in a single image and was exploited both for image authentication and geo-location estimation \cite{b11}.

Apart from authenticating multimedia content, ENF can be exploited for audio and video synchronization \cite{b13}. The region of multimedia recordings can be determined by machine learning systems without the need of a reference signal \cite{b14}. Edit detection in multimedia recordings can be achieved by exploiting ENF variations, as done in \cite{b15}.

Moreover, spectral distances as linear predictors and phase change analysis are exploited in order to justify whether multimedia content has been tampered \cite{b16}. Timestamp verification and tampering detection are handled by employing absolute-error-maps in a couple of methods for exhaustive point search and image erosion, as presented in \cite{b18}. Classification schemes using machine learning algorithms for tampering detection employing an Estimation of Signal Parameters via Rotational Invariance Techniques (ESPRIT)-based method are presented in  \cite{b20}. A systematic study on factors affecting ENF in audio recordings is presented in \cite{b21}.

In this study, motivated by \cite{b22,b23}, we introduce a Capon-based approach for ENF estimation. The proposed scheme employs a temporal window and exploits the Toeplitz structure of the covariance matrix. Together with Krylov matrices and Gohberg-Semencul (GS) factorization derives a fast and effective approach for ENF estimation. A Parzen window is employed as a temporal window, which is shown to yield accurate ENF estimation. Having seen the boost that ENF estimation gains due to the proper temporal window selection, we examine various windows of different lengths. In parallel, we explore the effect of window selection on the ENF extraction, using  Short-Time Fourier Transform (STFT).

\section{WINDOW SELECTION AND ESTIMATION PROCEDURE}

Window selection was not thoroughly investigated within ENF estimation. The rectangular window has been used exclusively as a temporal window \cite{b3}. Temporal windowing denotes the multiplication of the time-series with a window prior to spectral analysis. On the contrary, a lag window denotes the multiplication of the sequence of autocovariances for various lags with a window \cite{b30}. In this study, we employ temporal windowing. It is shown through extensive experiments that the selection of window function pays off. Proper window selection is able to provide finer spectral resolution and boost the  accuracy of frequency estimation. For example, the $N$-point Parzen window \cite{b24, b25} is defined as:

\begin{equation}
\resizebox{0.91\hsize}{!}{$%
w(n)= \begin{cases}
1 - 6\bigg(\dfrac{|n|}{N/2}\bigg)^2 + 6\bigg(\dfrac{|n|}{N/2}\bigg)^3   & 0\leq |n| \leq \big(N-1\big)/4\\
2\bigg(1-\dfrac{|n|}{N/2}\bigg)^3 & \big(N-1\big)/4 \leq |n| \leq \big(N-1\big)/2.
\end{cases}
$}
\end{equation}
Along with the Parzen window, we also employ Kaizer, Hamming, and rectangular windows \cite{b26}.

In order to estimate the ENF embedded in audio/power recordings, we followed the general scheme proposed in \cite{b27} and the parametrization suggested in \cite{b8}. The first step includes the proper filtering of the raw signal. A sharp zero-phase band-pass FIR filter with $C_1=1001$ and $C_2 = 4801$ coefficients was applied around the 3rd harmonic of the signal recorded from power mains and around the 2nd harmonic of the speech signal, respectively. A more detailed description of the dataset is presented in Sec. \ref{data}. In both cases, a tight band-pass frequency range of $0.1$\,Hz is employed. The frequency is estimated per frame. Between consecutive frames, there exists $1$ second shift, which is translated to $441$ samples for the downsampled sampling frequency of $441\,Hz$. Each frame is then multiplied by a temporal window. Afterwards, the maximum periodogram value of each frame, which corresponds to an approximate ENF estimation $\omega_{q_{\max}}$, is derived. In order to obtain a more precise estimation of ENF, we employ a quadratic interpolation, as done in \cite{b3,b28}. A quadratic model is fit to the logarithm of the estimated power spectrum \cite{b3}.% By using the two adjacent frequencies around $\omega_{\ell{\max}}$, a more accurate ENF estimate is obtained as  $\omega=\omega_{\ell_{\max}}+ \delta$, where

%\begin{equation}
%\delta=\frac{1}{2} \frac{\beta_{-1}-\beta_1}{\beta_{-1}-2\beta_0+\beta_1}( \omega_{\ell_{\max}+1}-\omega_{\ell_{\max}})
%\end{equation}
%and $\beta_q= \log \hat{\phi}_r(\omega_{\ell_{\max}+q})$,  $q=-1, 0, 1$.

After periodogram-based ENF estimation, in order to evaluate the results of the estimated frequencies, one employs the maximum correlation coefficient, as proposed in \cite{b29}. Using the notation introduced in \cite{b3}, let $\mathbf{f} = \left[f_1, f_2, \dotsc, f_K \right]^T$ be the estimated ENF signal at each second. Let also $\mathbf{g} = \left[ g_1, g_2, \dotsc, g_{\tilde{K}}\right]^T$ for $\tilde{K}>K$ be the reference ground truth ENF and $\mathbf{g}(l)=\left[ g_l, g_{l+1}, \dotsc, g_{l+K-1}\right]^T$ be a segment of $\mathbf{g}$ starting at $l$. One is seeking for

\begin{equation}
l_{opt}=\argmaxl_{l} c(l)
\end{equation}
where $l=1,2,\dots, \tilde{K}-K+1$ and $c(l)$ is the sample correlation coefficient between $\mathbf{f}$ and $\tilde{\mathbf{g}}(l)$ defined as:

\begin{equation}
c(l) = \frac{\mathbf{f}^T \tilde{\mathbf{g}}(l)}{\norm{\mathbf{f}}_2 \norm{\tilde{\mathbf{g}}(l)}_2}.
\end{equation}
That is, ENF estimation is cast as  a maximum correlation matching between sequences $\mathbf{f}$ and $\mathbf{g}$. In the next section, we propose a data-driven approach for estimating $\mathbf{f}$ based on a fast Capon method. Moreover, in Section~\ref{res}, we study whether pairwise differences between the maximum correlation coefficient delivered by proposed fast Capon ENF estimation method employing temporal windows and that of other state-of-the-art ENF estimation methods are statistically significant. Another method for matching the extracted ENF to the ground
truth is the dynamic time warping (DTW) \cite{b8}.

\section{PROPOSED APPROACH }\label{proposed}

\subsection{The Capon method}

The periodogram can be interpreted as a filter bank approach, which uses a band-pass filter whose impulse response vector is given by the standard Fourier transform vector $\left[1, e^{-i\omega}, \ldots, e^{-i(N-1)\omega} \right]^T$. The Capon method, is another filter bank approach based on a data-dependent filter \cite{b30}:
\begin{equation}
\mathbf{h}=\frac{\hat{\mathbf{R}}^{-1} \mathbf{a}(\omega)}{\mathbf{a}^{*}(\omega) \, \hat{\mathbf{R}}^{-1} \, \mathbf{a}(\omega)}
\label{Caponh}
\end{equation}
where $\mathbf{a}(\omega)=\left[1, e^{-j\omega}, \ldots, e^{-j \,m \, \omega}\right]^{*}$ and $[\cdot]^{*}$ denotes conjugate transposition.
In (\ref{Caponh}), $\hat{\mathbf{R}}$ is an estimate of the auto-covariance matrix
\begin{equation}
\resizebox{.95\hsize}{!}{$
\hat{\mathbf{R}}=\frac{1}{N-m} \sum^N_{t=m+1} \begin{bmatrix}\tilde{y}(t) \\ \vdots\\ \tilde{y}(t-m)\end{bmatrix} \begin{bmatrix}\tilde{y}^{*}(t), \dotsc, \tilde{y}^{*}(t-m)\end{bmatrix}$.}
\end{equation}
where $\tilde{y}(t)= w(t-n)y(t) $. Here, $m=10$ and $N=L\, F_s$, where $L$ is the frame length in sec.  The Capon spectral estimate is given by:
\begin{equation}
\hat{\phi}(\omega) =\frac{m+1}{\mathbf{a}^{*}(\omega) \, \hat{\mathbf{R}}^{-1}\, \mathbf{a}(\omega)}.
\label{eq:Capon}
\end{equation}
Eq.~\eqref{eq:Capon} is computed for dense frequency samples $\omega_{q}=\frac{2 \pi q}{Q}$, $q=0,1, \ldots, Q-1$ with $Q=4N$ every sec. 
ENF is estimated by the angular frequency sample $\omega_{q_{\max}}$ where the Power Spectral Density (PSD) of Eq.~\eqref{eq:Capon} attains a maximum  for $q \in [0, \frac{Q}{2}-1]$ and $f_k=\frac{\omega_{q_{\max}}}{2\pi} \, F_s$.
The Capon method has been found to be able to resolve fine details of PSD, making it a superior alternative of periodogram-based methods for ENF estimation.

 \subsection{Fast implementation}

The proposed approach exploits the Toeplitz structure of the covariance matrix $\mathbf{R}_{N}$ in order to reduce the computational complexity of the inversions included in the process of Capon spectral estimation. It is not limited to Toeplitz structures, but is expanded to low displacement rank matrices, where the displacement representation of any square matrix $\mathbf{A}$ is defined as \cite{b22}:
\begin{equation}
 \nabla_{\mathbf{D}_{N}, \mathbf{D}_{N}^{T}} \mathbf{A} = \mathbf{A} - \mathbf{D}_{N} \mathbf{A} \mathbf{D}_{N}^{T}
\end{equation}
with $\mathbf{D}_{N}$ being a lower triangular matrix, such as the lag-1 shift matrix \cite{b22}:

\[
  \mathbf{D}_{N} =
  \begin{bmatrix}
    0 & & & \\
    1&  \ddots  &  & \\
    &  \ddots & 0  & \\
    & &1  & 0
  \end{bmatrix}
\]

The core of this approach lies on the fast and efficient inversion of $\mathbf{R}_N^{-1}$, using the GS factorization. To do so, we exploit the Krylov matrix. Given an arbitrary vector $\mathbf{v}_{N}$ and a lower triangular matrix $\mathbf{D}_{N}$, Krylov matrix is defined as follows:
\begin{equation}
\mathcal{K}_{N}\big(\mathbf{v}_{N}, \mathbf{D}_{N}\big)  = \big[ \mathbf{v}_{N}, \mathbf{D}_{N} \mathbf{v}_{N}, \cdots, \mathbf{D}_{N}^{N-1} \mathbf{v}_{N}  \big]
\end{equation}
Exploiting the Krylov matrix and the unit lower shift matrix $\mathbf{D}_{N}$, the inverse of $\mathbf{R}_{N}$ is formulated as \cite{b22,b30}:

\begin{equation}
\begin{split}
\mathbf{R}_{N}^{-1} &= \mathcal{K}_{N}\big(\boldsymbol{\gamma}_{N}, \mathbf{D}_{N}\big) \mathcal{K}_{N}^{*} \big(\boldsymbol{\gamma}_{N}, \mathbf{D}_{N}\big)\\  &- \mathcal{K}_{N}\big(\boldsymbol{\delta}_{N}, \mathbf{D}_{N}\big) \mathcal{K}_{N}^{*} \big(\boldsymbol{\delta}_{N}, \mathbf{D}_{N}\big)
\end{split}
\end{equation}
where

\begin{equation}
\boldsymbol{\gamma}_{N} = \dfrac{1}{\sqrt{\alpha_{N-1}}} \begin{bmatrix}
   1 \\

    \mathbf{w}_{N-1}
\end{bmatrix}
\end{equation}

\begin{equation}
\boldsymbol{\delta}_{N} = \dfrac{1}{\sqrt{\alpha_{N-1}}} \begin{bmatrix}
   0 \\

   \mathbf{\breve{w}}_{N-1}^{*}
\end{bmatrix}
\end{equation}
where the vector $\mathbf{\breve{w}}_{N-1}^{*}$ is the conjugate transpose version of $\mathbf{w}_{N-1}$ with the order of its elements reversed.

The parameters $\boldsymbol{\gamma}_{N}$ and $\boldsymbol{\delta}_{N}$ can be computed  through a system of linear equations, using the Levinson-Durbin algorithm at a cost of $N^{2}$ operations. The covariance matrix $\mathbf{R}_{N}$ can be partitioned as:

\begin{equation}
\mathbf{R}_{N} =  \begin{bmatrix}
   \rho_{0} &\boldsymbol{\rho}^{*}_{N-1} \\

    \boldsymbol{\rho}_{N-1} & \mathbf{R}_{N-1}
\end{bmatrix}
\end{equation}
where $\rho_{0}$ is the element $(1,1)$ of $\mathbf{R}_{N}$, $\boldsymbol{\rho}_{N-1}$ is a column vector with entries the rest of the elements of the first column of $\mathbf{R}_{N}$, while $\boldsymbol{\rho}^{*}_{N-1}$ represents the conjugate transpose row vector of $\boldsymbol{\rho}_{N-1}$. The parameters $\boldsymbol{\gamma}_{N}$ and $\boldsymbol{\delta}_{N}$ are computed by solving the following system for $ \mathbf{w}_{N-1}$ \cite{b23, b31}:

\begin{equation}
\mathbf{R}_{N-1} \mathbf{w}_{N-1} = -\boldsymbol{\rho}_{N-1}
\end{equation}
and calculating $\alpha_{N-1}$ through $\alpha_{N-1} = \rho_{0} + \boldsymbol{\rho}^{*}_{N-2}\mathbf{w}_{N-2}$.

In order to efficiently calculate the denumerator of Eq. \eqref{eq:Capon}, we can alternatively perform zero-padded FFT on the coefficients $x_{i}$ of a univariate polynomial \cite{b22}:

\begin{equation}
\phi_{den}(\omega_{q}) = \mathbf{a}^{*}(\omega_{q}) \, \hat{\mathbf{R}}^{-1}\, \mathbf{a}(\omega_{q}) = \sum_{i = -N+1}^{N-1} x_{i} \; e^{j \frac{2 \pi \,q}{Q} \, i}
\end{equation}
where $q = 0,1,\cdots,\,Q-1$.

\section{EXPERIMENTAL EVALUATION}
\subsection{Datasets}\label{data}

Two datasets are employed in order to evaluate the proposed approach and compare it with other state-of-the-art methods. The first dataset, namely Data 1, was recorded by connecting an electric outlet directly to the internal sound card of a desktop computer, while the second one, namely Data 2, comprises of a speech recording captured by the internal microphone of a laptop computer. Both datasets were also used in \cite{b3,b4,b7,b8}. The original datasets were sampled at $44.1\,kHz$ using $16$ bits per sample. Afterwards, the initial recordings were downsampled at $441\,Hz$, using proper anti-aliasing filtering. Apart from the fundamental frequency of $60\,Hz$, the second and the third harmonics were also maintained to perform experiments. Regarding the first dataset, only the third harmonic was used, because it provides the best results compared to the other two. In the speech recording, the second harmonic was used, because the other two suffered from extremely low SNR. The third recording was the reference ground truth ENF $\mathbf{g}$ captured by a Frequency Disturbance Recorder. The extracted ENF signal was compared against the ground truth.

\subsection{Results} \label{res}

In order to evaluate the proposed approach, we employed the two datasets mentioned in Sec. \ref{data} and compared it against state-of-the-art approaches. A $30$ min long recording of each dataset was used in the experiments. Regarding Data 1, four different frame lengths of $1$, $5$, $10$, and $20$ sec  were used for ENF estimation. As stated in Table \ref{tab:cord1}, by applying the procedure in Sec. \ref{proposed} a correlation coefficient of $0.9990$ is obtained, when a frame length of $20$ seconds is employed. This value exceeds the state-of-the-art linear prediction estimation presented in \cite{b7}, which reaches $0.9984$ even though their value is not purely from the linear prediction method due to the fact that there has been made a denoising procedure before. Moreover, the aforementioned value of $0.9990$ overcomes the Maximum Likelihood (ML) method and the Welch one, which was properly parametrized in \cite{b8}. An interesting fact regarding our proposed method lies in the frame length. When shorter frame lengths of $5$ and $10$ sec are employed, the accuracy gets higher and reaches $0.9991$, while for the other methods accuracy drops as the frame length gets shorter. Due to the need for fast and accurate ENF estimation in real world applications, it is seen that even when an $1$ second frame length is adopted, an accuracy of $0.9990$ is obtained. Despite the very short frame length, there is a high resolution enabling the accurate estimation of the ENF. For all other methods, accuracy drops below $0.99$ within this setup and reaches $0.8255$ when weighted spectrogram is used.

In addition to the comparisons between the different approaches for ENF estimation, a systematic study was carried out in order to examine the impact the different windows have on ENF estimation. Four different windows along with four different frame lengths were employed, as shown in Table \ref{tab:corwind1}. As stated before, the Parzen window yields the highest accuracy among approaches and is not affected by the frame length at all. This makes it the best choice in ENF estimation applied to a recording from power mains. The performance of the Hamming window is similar, which makes it a good alternative to Parzen window. On the contrary, Kaiser and rectangular windows yield remarkable results only when a larger frame length is employed. Even when a $10$ seconds frame is used, Kaiser and rectangular windows are not able to provide accurate ENF estimation. Therefore, the choice of the window is not a trivial task and impacts accurate ENF estimation.

\begin{figure}[!t]
\centering
\scalebox{0.67}{\includegraphics{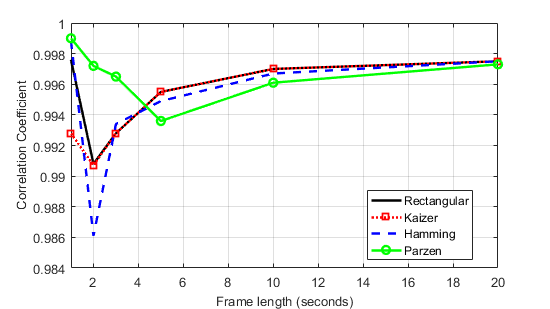}}
\caption{STFT using different windows for ENF estimation of Data 1.}
\label{figspectro}
\end{figure}

%\begin{figure}[!t]
%\caption{Different windows for ENF estimation of Data 2.}
%\centering
%\scalebox{0.37}{\includegraphics{data_2_windows.png}}
%\label{figmdata2win}
%\end{figure}

Employing the proper window is essential in every approach chosen for ENF estimation. In order to demonstrate the crucial role of window selection, we present the accuracy obtained using the trivial method of the STFT for various windows. As shown in Fig. \ref{figspectro}, Parzen window yields highly accurate results even when an $1$ second frame length is employed. This accuracy approaches $0.9990$, which means that the STFT approach with proper temporal window can outperform state-of-the-art methods. It is also evident that even though high accuracy is achieved when long frame lengths are used, the selection is very important when shorter ones are employed. Additionally, it is demonstrated that longer frame lengths do not necessarily imply better accuracy in terms of correlation coefficient.

\begin{table}[!htb]
  \centering
  \caption{Correlation coefficient for various frame lengths - Data 1}
  \label{tab:cord1}

\scalebox{0.82}{
  \begin{tabular}{l c c c c}\hline
     Frame length (in sec)            & 1 &5 & 10 & 20 \\ \hline
  \textbf{Proposed with Parzen window}   & $0.9990$ & $0.9991$ & $0.9991$ & $0.9990$\\
  \textbf{ML \cite{b4}} & $0.8826$ & $0.9852$ & $0.9953$ & $ 0.9977$ \\
    \textbf{Linear Prediction \cite{b7}}    & $ 0.9651$ & $0.9959$ & $0.9976$ & $0.9984$\\
    \textbf{Welch\cite{b8}}       & $0.9847 $ & $0.9989 $ & $0.9989 $ & $0.9983$ \\
    \textbf{Weighted Spectrogram \cite{b7}}       & $0.8255$ & $0.9873$ & $0.9944$ & $0.9966$ \\

 \hline

  \end{tabular}
}
\end{table}

\begin{table}[!htb]
  \centering
  \caption{Correlation coefficient for various windows - Data 1}
  \label{tab:corwind1}

\scalebox{0.95}{
  \begin{tabular}{l c c c c}\hline
     Frame length (in sec)            & 1 &5 & 10 & 20 \\ \hline
  \textbf{Parzen}   & $0.9990$ & $0.9991$ & $0.9991$ & $0.9990$\\
  \textbf{Hamming} & $0.9989$ & $0.9991$ & $0.9990$ & $ 0.9988$ \\
    \textbf{Kaizer}    & $ 0.0086$ & $0.0495$ & $0.0438$ & $0.9976$\\
    \textbf{Rectangular}       & $0.0047$ & $0.0798$ & $0.0689$ & $0.9975$ \\

 \hline

  \end{tabular}
}
\end{table}

The second dataset (Data 2) comprises of speech recordings in which interference exist and suffers from low SNR. In Data 2, we studied the second harmonic, where a higher SNR permits us to obtain reliable results. Our proposed approach (Sec. \ref{proposed}) is extremely fast due to the fact that it exploits Krylov matrices and the Toeplitz structure of the covariance matrix. It also provides high frequency resolution, yielding an accuracy of $0.9351$ in terms of correlation coefficient. This result is obtained using a $33$ sec frame length and a rectangular window. Our approach outperforms the existing ML approach and the high resolution MUSIC method, as demonstrated in Table \ref{tab:cord233}. It is lagging behind the pure linear prediction method without the additional denoising procedure with respect to the correlation coefficient for about $0.0015$, but is still meant to be an efficient alternative taking into account the fact that linear prediction is an iterative approach, which includes large matrix inversions in each iteration. When speaking of large datasets, like the one we discuss in this study, linear prediction is much slower than the proposed approach. Consequently, the trade-off between accuracy and time complexity of our approach constitutes a useful tool in ENF estimation, no matter what the nature of the recordings and the duration are.

A systematic study was also carried out in order to examine the impact of the window on the speech recordings of Data 2. Four windows were employed, as done for Data 1, with four different frame lengths, as presented in Table \ref{tab:corwind2}. In speech recordings, increasing frame length provides better results at the expense of time requirements. It is also evident that for very short frame lengths (i.e., $5$ sec) accuracy is deteriorating rapidly. However, the window, which is going to be selected, plays a key role in the final accuracy of ENF estimation. Although all four choices provide acceptable results in terms of accuracy, it is the rectangular window, which outperforms its competitors now.

In order to determine whether the correlation coefficient of the proposed method is significantly different from  that of other methods ($H_{1}$: $c_1 \neq c_2$), hypothesis testing was applied. Fisher transformation, $z=0.5  \ln \frac{1+c}{1-c}$ was employed for each pair of correlation coefficients under examination \cite{Papoulis}. For significance level $95\%$, the test statistic $q = \sqrt{n-3}\,(z_{1}-z_{2})$ was outside the region $-1.96 < q < 1.96$, where $n=1800$. Thus, the null hypothesis was rejected for every pair of comparisons. Accordingly, the differences between the correlation coefficients were significant at confidence level of $95\%$.

\begin{table}[!htb]
  \centering
  \caption{Correlation coefficient for various frame lengths - Data 2}
  \label{tab:cord233}

  \begin{tabularx}{\columnwidth}{l c c}\hline
      Frame length (in sec)  & 10 & 33   \\ \hline
     \textbf{Proposed with rectangular window} & $0.8663$ & $0.9351$\\
     \textbf{ML \cite{b4}}& $0.9059$ & $0.9319$ \\
    \textbf{Linear Prediction \cite{b7}} &  $0.9213$ & $ 0.9366$\\
     \textbf{MUSIC \cite{b8}}   &  $ 0.9087$  & $0.9318$ \\
    \textbf{Weighted Spectrogram \cite{b7}} &   $ 0.8787$   & $0.9125$ \\

 \hline

  \end{tabularx}

\end{table}

\begin{table}[!htb]
  \centering
  \caption{Correlation coefficient for various windows - Data 2}
  \label{tab:corwind2}

\scalebox{0.95}{
  \begin{tabular}{l c c c c}\hline
     Frame length (in sec)            &5 & 10 & 20 & 33 \\ \hline
  \textbf{Parzen}   & $0.7063$ & $0.7773$ & $0.8413$ & $0.8785$\\
  \textbf{Hamming} & $0.7453$ & $0.8128$ & $0.8703$ & $ 0.8987$ \\
    \textbf{Kaizer}    & $ 0.8081$ & $0.8663$ & $0.9035$ & $0.9228$\\
    \textbf{Rectangular}       & $0.8092$ & $0.8663$ & $0.9036$ & $0.9351$ \\

 \hline

  \end{tabular}
}
\end{table}

\section{CONCLUSION}\label{conc}
A novel approach for ENF estimation based on Capon method with temporal windowing has been discussed. Taking advantage of the Toeplitz structure of the covariance matrices and exploiting Krylov matrices, a fast and efficient approach has been developed, which yields higher accuracy compared to the state-of-the-art methods in power recordings. Furthermore, the aforementioned approach works fast and properly, and provides state-of-the-art results even when  very short frame lengths are employed. Accordingly, an excellent trade-off between speed and accuracy is offered by the proposed method, which is of crucial importance in forensic applications. Moreover, through extensive experiments, we have shown that the choice of the window is not a trivial task and methods such as the STFT can achieve very good results, even when a very short window is adopted. Exploring the potential implementation of this algorithm for estimating ENF on the fly from single images in real world forensics applications could be a topic for future research.

%\newpage
%\newpage
\bibliographystyle{IEEEbib}

\bibliography{bibliog}

\end{document}